\documentclass[runningheads]{svmult}

\usepackage{makeidx}   
\usepackage{graphicx}  
\usepackage{subeqnar}  
\usepackage{multicol}  
\usepackage{physprbb}  
\makeindex             

\newcommand{\lya}{Ly-$\alpha$\ }
\newcommand{\dndz}{$dN/dz$}
\newcommand{\ggh}{\Gamma_{\rm HI}}
\newcommand{\nhi}{$N_{HI}$}
\newcommand{\nh}{N_{HI}}
\newcommand{\cm}{\rm cm} 
\newcommand{\magcir}{\ \raise -2.truept\hbox{\rlap{\hbox{$\sim$}}\raise5.truept
 	\hbox{$>$}\ }}		
\newcommand{\mincir}{\ \raise -2.truept\hbox{\rlap{\hbox{$\sim$}}\raise5.truept
	\hbox{$<$}\ }}	
\begin{document}
\title*{Exploring the intergalactic medium with VLT/UVES}
\toctitle{Exploring the intergalactic medium with VLT/UVES}
%
%
\titlerunning{Exploring the IGM with VLT/UVES}
%
\author{Stefano Cristiani\inst{1,2}
\and Simone Bianchi\inst{3}
\and Sandro D'Odorico\inst{3}
\and Tae-Sun Kim\inst{3}
}
\authorrunning{Stefano Cristiani et al.}
%
%
\institute{ST European Coordinating Facility, K.-Schwarzschild-Str. 2,
D-85748 Garching
\and
Osservatorio Astronomico di Trieste, via Tiepolo 11, I-34131 Trieste
\and 
European Southern Observatory, K.-Schwarzschild-Str. 2, D-85748 Garching
}
\maketitle              

\begin{abstract}
The remarkable efficiency of the UVES spectrograph at the VLT has made
it possible to push high-resolution, high-S/N ground observations of
the \lya forest down to $z \sim 1.5$, gaining new insight into the
physical conditions of the intergalactic medium and its evolution over
more than 90\% of the cosmic time.  The universal expansion, the UV
ionizing background and the gravitational condensation of structures
are the driving factors shaping the number density and the column
density distribution of the absorbers. A (limited) contribution of UV
photons produced by galaxies is found to be important to reproduce the
observed evolutionary pattern at very high and low redshift.  The
Lyman forest contains most of the baryons, at least at $z>1.5$, and
acts as a reservoir for galaxy formation. The typical Doppler
parameter at a fixed column density is measured to slightly increase
with decreasing redshift, but the inferred temperature at the mean
density is increasing with redshift. The signatures of HeII
reionization and feedback from the formation of galactic structures
have possibly been detected in the Lyman forest.
\end{abstract}
\section{The Observations}
A sample of 8 QSOs with $1.7 < z_{\rm em} < 3.7$ has been observed
\cite{kim01,kim02} with VLT/UVES at a typical resolution $45000$ and
S/N $\sim 40-50$.  Thanks to the two-arm design of the spectrograph
\cite{UVES}, a high efficiency has been achieved in the whole optical
range, from the atmospheric cutoff to $1 \mu m$, which translates
immediately in the possibility of obtaining new results on the Lyman
forest, especially at $z \mincir 2.5$.  The data have been reduced
with the UVES pipeline \cite{UVESpip} - an non-negligible factor in
maximizing the scientific output per unit time - and analyzed with the
VPFIT package \cite{VPFIT}.

\section{The number density and column density evolution of \lya lines}
\begin{figure}[t]
\begin{center}
\includegraphics[width=1.\textwidth]{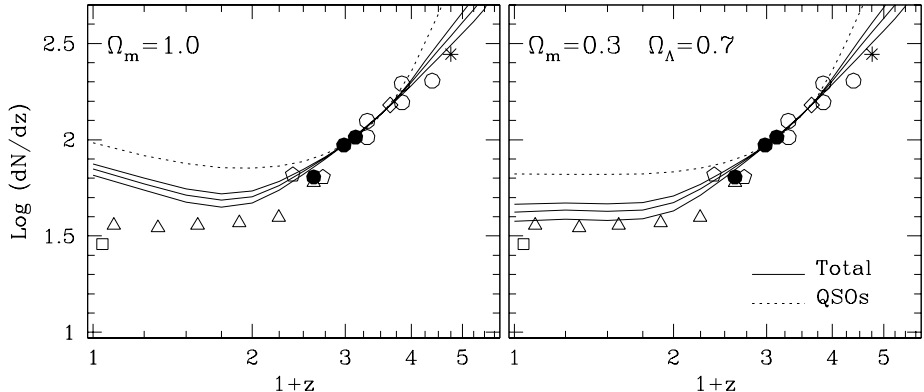}
\end{center}
\caption[]{ Number density evolution of the Ly$\alpha$ forest with 
$\nh =10^{13.64-16} \;\mathrm{cm^{-2}}$.
Dotted lines refer to the evolution compatible with 
an ionising UV background due only to QSOs. Solid lines show the 
expected evolution 
when both QSOs and galaxies contribute to the background, for models 
with $f_\mathrm{esc}$=0.05 (upper line), 0.1 and 0.4 (lower line). 
Data points come from several observations in the literature, 
as given by \cite{kim01}. 
The modelled evolution has been normalized to the observed evolution in 
the redshift range $2<z<3$.
\cite{bianchi01}
}
\label{dndz}
\end{figure}

The swift increase of the number of absorptions (and the average
opacity) with increasing
redshift is the most impressive property of the \lya forest.
Fig.~\ref{dndz} shows the number density evolution of the
\lya lines \cite{kim01,kim02} in the column density interval
\footnote{This range in \nhi\ has been chosen to allow a
comparison with the HST Key-Programme sample at $z < 1.5$ \cite{weymann98}
for which a threshold in equivalent width of 0.24 \AA\/ was adopted.}
$N_{HI} = 10^{13.64 - 16} \ {\rm cm}^{-2}$.
The maximum-likelihood fit to the data at $z > 1.5$ with the customary
power-law parameterization provides $N (z) = N_{0}
(1+z)^{\gamma} = (6.5 \pm 3.8)\,(1+z)^{2.4 \pm 0.2}$. 
The UVES observations imply that the turn-off in the
evolution does occur at $z \sim 1$, not at $z \sim 2$ as
previously suggested.  

While the opacity is varying so fast, the column density distribution
stays almost unchanged.
%
%
The differential density distribution function measured by UVES
\cite{kim01,kim02}, that is the number of lines per unit redshift path
and per unit \nhi\ as a function of \nhi, basically follows a
power-law $f( \nh ) \propto \nh^{-1.5}$ extending over 10 orders of
magnitude with little, but significant deviations: the slope $\beta$
of the power-law in the range $14 \mincir \log \nh \mincir 16 $
goes from about $-1.5$ at $<z> = 3.75$ to $-1.7$ at $z < 2.4$.
Recent HST STIS data \cite{dave01} confirm that this trend continues
at lower redshift, measuring a $\beta$ of $-2.0$ at $z<0.3$.
\section{The evolution of the \lya forest and the ionizing background}
The evolution of the $N(z)$ is governed by two main factors: the
Hubble expansion and the metagalactic UV background (UVB).
At high $z$ both the expansion, which decreases the density
and tends to increase the ionization,
and the UVB, which is increasing or non-decreasing 
with decreasing redshift, work in the same direction and
cause a steep evolution of the number of lines.
At low $z$, the UVB starts to decrease with decreasing
redshift, due to the reduced number and intensity of the ionizing sources, 
counteracting the Hubble expansion. As a result the evolution 
of the number of lines slows down.

Up to date numerical simulations \cite{theuns98} have been remarkably
successful in qualitatively reproducing the observed evolution,
however they predict the break in the \dndz\ power-law at a
redshift $z \sim 1.8$ that appears too high in the light of the new
UVES results. This suggests that the UVB implemented in the
simulations may not be the correct one: it was thought that at low
redshift QSOs are the main source of ionizing photons, and, since
their space density drops below $z\sim 2$, so does the UVB.
However, galaxies can produce a conspicuous ionizing flux too, possibly
more significant than it was thought\cite{steidel01}. 
The galaxy contribution can keep the
UVB relatively high until at $z \sim 1$ the global star
formation rate in the Universe quickly decreases, determining the
qualitative change in the number density of lines.

Under relatively general assumptions, it is possible to
relate the observed number of lines above a given threshold in column
density or equivalent width to the
expansion, the UVB, the distribution in column density of 
the absorbers and the cosmology \cite{dave99}:
\begin{equation}
\left(dN \over dz\right)_{>N_{HI,\rm lim}} = 
C \left[(1+z)^5 \ggh^{-1}(z)\right]^{\beta-1} H^{-1}(z),
\label{eq:dndz}
\end{equation}
where $\ggh$ is the photoionization rate and  $\beta$ the power-law
index of the \nhi\ distribution.

To estimate $\ggh$ we have investigated the contribution of
galaxies to the UVB\cite{bianchi01}, exploring three values for the
fraction of ionizing photons that can escape the galaxy interstellar
medium, $f_{esc} = 0.05, 0.1$ and $0.4$ (the latter value corresponds
to the Lyman-continuum flux detected by \cite{steidel01} in the
composite spectrum of 29 Lyman-break galaxies).
Measurements of the UVB based on the proximity effect at high-$z$ and on
the ${\rm H}\alpha$ emission in high-latitude galactic clouds at
low-$z$ provide an upper limit on $f_{esc} \mincir 0.1$, consistent with
recent results on individual galaxies both at low-$z$
\cite{deharveng01,heckman01} and at $z\sim3$ \cite{giallongo01}.
Introducing a contribution of galaxies to the UVB,
the break in the \lya \dndz\ can be better
reproduced than with a pure QSO contribution \cite{bianchi01}. 
The agreement improves
considerably also at $z\magcir 3$. Besides, models with $\Omega_{\Lambda}
=0.7, \Omega_{M}=0.3$ describe the flat evolution of the absorbers
much better than $\Omega_{M}=1$.
\begin{figure}[t]
\begin{center}
\includegraphics[width=1.05\textwidth]{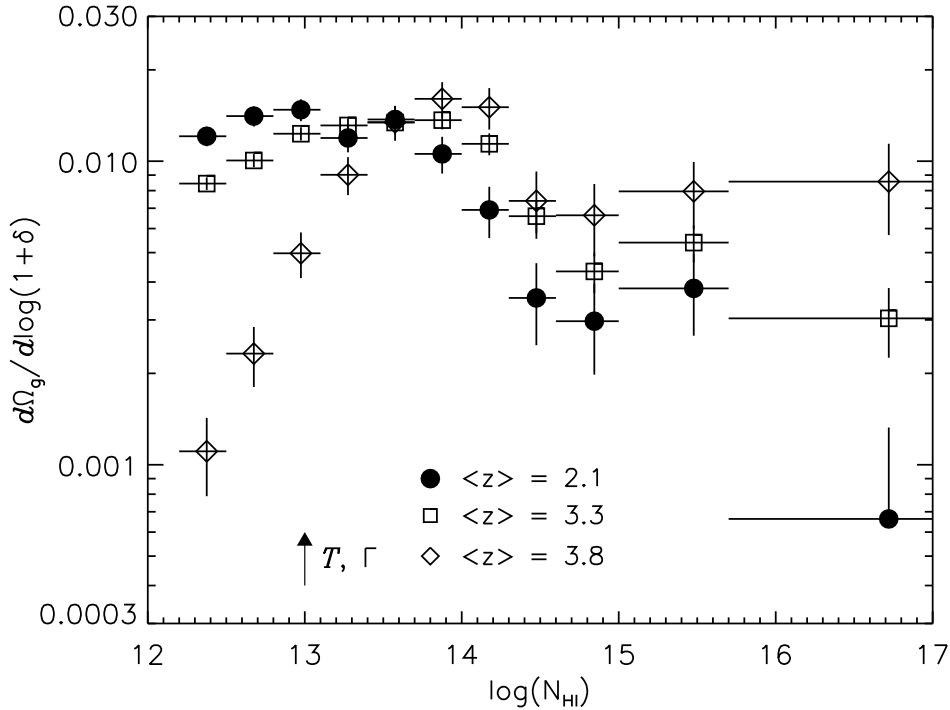}
\end{center}
\caption[]{The differential mass density distribution of the \lya
forest as a function of \nhi. The arrow indicates the direction
towards which the points move if the temperature or the ionization
rate increase \cite{kim02},}
\label{NHImass}
\end{figure}
A consistency check is provided by the evolution of the lower column
density lines. For $\log \nh \mincir 14$ the \nhi\ distribution
follows a flatter slope $\beta$,
and according to Eq.~\ref{eq:dndz} this translates directly
into a slower evolutionary rate, which is consistent
with the UVES observations\cite{kim01}: $\dndz_{(13.1<\nh<14)} \propto
(1+z)^{1.2\pm0.2}$. Another diagnostic can be
derived from the spectral shape of the UVB and its influence
on the intensity ratios of metal lines \cite{savaglio97,songaila98}.
\section{Mapping the column density distribution into the mass
distribution of the gas}
It is instructive to transform the observed
column density distribution in the mass distribution of the
photoionized gas (Fig.~\ref{NHImass}) and 
interpret it, following Schaye\cite{schaye01}, 
as a function of the matter density contrast:
1) the flattening at $\log N_{HI}\mincir 13.5$ is partly due to line crowding
and partly to the turnover of the density distribution below the mean
density;
2) the steepening at $\log \nh \magcir 14$, with a deficiency of lines
that becomes more and more evident at lower z, reflects the fall-off
in the density distribution due to the onset of rapid, non-linear
collapse
3) the flattening at $N_{HI} \magcir 10^{16}~\cm^{-2}$ can be attributed to
the flattening of the density distribution at density contrast 
$\magcir 10^2$ due to the virialization of collapsed matter.
The differential mass density distribution has a sort of universal
form when plotted as a function of the density contrast. A given
density contrast, however, corresponds to lower and lower column
densities with decreasing redshift, and this causes a translation of
the mass density distribution (Fig.~\ref{NHImass}) towards the left
with decreasing redshift, which explains the steepening of the slope
$\beta$ reported in Sect.2.
Hydrodynamical simulations successfully reproduce this behaviour,
indicating that the derived matter distribution is indeed
consistent with what would be expected from gravitational instability.
\section{The cosmic baryon density}
The amount of baryons required in a given cosmological scenario to
produce the observed opacity of the Lyman forest can be computed
\cite{weinberg97} under general assumptions.
A lower-bound to the cosmic baryon density can be derived from the
mean \lya flux decrement, ${\overline{D}}$,\cite{oke82} and/or
from the distribution of the \lya optical depths.
The limits derived from the 
effective optical depths measured in the UVES spectra at $1.5<z<4$
are reported in Tab.~\ref{tab:omega}.
They are consistent with the BBN value for a low D/H primordial abundance.
Most of the baryons reside in the Lyman
forest at $1.5<z<4$ with little change in the contribution to
$\Omega$ as a function of $z$.
Conversely, given the observed opacity, a higher UVB requires a
higher  $\Omega_b$. As pointed out by \cite{haehnelt01}, an escape
fraction as large as 0.4, as measured by \cite{steidel01}, would result
in an $\Omega_b \sim 0.06$ in conflict either with the primordial D/H 
abundance or in general with the BBN or with the \lya opacity measurements.
\begin{table}
\caption{Lower limits to $\Omega_b h^{1.5}$ derived from the UVES
spectra  at $1.5<z<4$
(for a universe with $\Omega_m=0.3, \Omega_{\Lambda}=0.7$)}
\begin{center}
\renewcommand{\arraystretch}{1.4}
\setlength\tabcolsep{5pt}
\begin{tabular}{|c|c|c|}
\hline\noalign{}
UVB  & ${\mathrm T} = 2 \cdot 10^4$ K &  ${\mathrm T} = 6 \cdot 10^3$ 
K \\
\hline\noalign{}
QSOs & 0.017 & 0.011 \\
QSOs + GALs & 0.028 & 0.018 \\
\noalign{\vskip -2mm}
(${\mathrm f}_{esc} = 0.1$)& & \\
\hline\noalign{}
\end{tabular}
\end{center}
\label{tab:omega}
\vskip -5mm
\end{table}
\section{The temperature of the IGM}
If the \lya forest is in thermal equilibrium with the metagalactic UV
background, the line width of the absorption lines, described by the
$b$ parameter of the Voigt profile, is directly related to the gas
temperature of the absorbing medium determined by the balance between
adiabatic cooling and photo-heating: $b = \sqrt{ 2 k T/ m_{ion}}$.
Additional sources of broadening exist, such as the
differential Hubble flow across the absorbers, peculiar motions, Jeans
smoothing.  However, there is a lower limit to the line widths, set by
the temperature of the gas, that is in principle measurable.
In practice the situation is slightly more complex
because for a photoionized gas there is a temperature-density
relation, the so-called equation of state: $T=T_{0} \,
(1+\delta_{b})^{\gamma_{T}-1}$, where $T$ is the gas temperature,
$T_{0}$ is the gas temperature at the mean gas density, $\delta_{b}$
is the baryon over-density,
$(\rho_{b}-\overline{\rho}_{b})/\overline{\rho}_{b}$ and $\gamma_{T}$
is a constant which depends on the ionization history.  
The equation of state translates into a lower cutoff
$b_c(N_{HI})$ in the $N_{HI}$--$b$ distribution.

The observed cut-off Doppler parameter at a fixed column density of 
$\log_{\nh}=13.6$, $b_c(13.6)$ 
is measured to increase with decreasing redshift, while
the slope of the cutoff does not change significantly.
A typical value of
$b_c \sim $18 km s$^{-1}$ at $1.5 <
z < 4$  corresponds to a reference temperature of $2 \cdot 10^{4}$K.
This does not mean that the temperature of the IGM increases with
decreasing $z$: on the contrary, taking into account the equation of
state and the fact that a given column density corresponds to higher and
higher over-densities with decreasing redshift, it turns out that the
temperature at the mean density is actually decreasing with
decreasing $z$.
Furthermore, evidence has been found \cite{theuns02} for a general
increase of the temperature around the redshift $z=3.3\pm0.15$,
attributed to the reionization of HeII.  Temperature and ionization
fluctuations are also expected due to feedback processes from the
formation of galactic structures \cite{theuns01} and might have been
observed in correspondence of some voids in the \lya forest
\cite{kim01}.

%

\end{document}